# Deep learning for drug repurposing: methods, databases, and applications

**Authors:**

| |
|---|
| **Xiaoqin Pan**<br>ORCID ID: 0000-0002-2083-338X<br>Affiliation: School of Computer Science and Engineering, Hunan University<br>Email: pxq123@hnu.edu.cn |
| **Xuan Lin**<br>ORCID ID: 0000-0001-7216-1821<br>Affiliation: School of Computer Science, Xiangtan University<br>Email: jack_lin@xtu.edu.cn |
| **Dongsheng Cao**<br>ORCID ID: 0000-0003-3604-3785<br>Affiliation: Xiangya School of Pharmaceutical Sciences, Central South University<br>Email: oriental-cds@163.com |
| **Xiangxiang Zeng**<br>ORCID ID: 0000-0003-1081-7658<br>Affiliation: School of Computer Science and Engineering, Hunan University<br>Email: xzeng@hnu.edu.cn |
| **Philip S. Yu**<br>ORCID ID: 0000-0002-3491-5968<br>Affiliation: Department of Computer Science, University of Illinois at Chicago<br>Email: psyu@uic.edu |
| **Lifang He**<br>ORCID ID: 0000-0001-7810-9071<br>Affiliation: Department of Computer Science and Engineering, Lehigh University<br>Email: lih319@lehigh.edu |
| **Ruth Nussinov**<br>ORCID ID: 0000-0002-8115-6415<br>Affiliation: Computational Structural Biology Section, Basic Science Program, Frederick National Laboratory for Cancer Research, National Cancer Institute at Frederick, Frederick, MD 21702, USA<br>Department of Human Molecular Genetics and Biochemistry, Sackler School of Medicine, Tel Aviv University, Tel Aviv 69978, Israel<br>Email: nussinor@mail.nih.gov |
| **Feixiong Cheng**<br>ORCID ID: 0000-0002-1736-2847<br>Affiliation: Genomic Medicine Institute, Lerner Research Institute, Cleveland Clinic, Cleveland, OH 44195, USA<br>Department of Molecular Medicine, Cleveland Clinic Lerner College of Medicine, Case Western Reserve University, Cleveland, OH 44195, USA.<br>Case Comprehensive Cancer Center, Case Western Reserve University School of Medicine, Cleveland, OH 44106, USA<br>Email: chengf@ccf.org |

## Abstract

Drug development is time-consuming and expensive. Repurposing existing drugs for new therapies is an attractive solution that accelerates drug development at reduced experimental costs, specifically for Coronavirus Disease 2019 (COVID-19), an infectious disease caused by severe acute respiratory syndrome coronavirus 2 (SARS-CoV-2). However, comprehensively obtaining and productively integrating available knowledge and big biomedical data to effectively advance deep learning models is still challenging for drug repurposing in other complex diseases. In this review, we introduce guidelines on how to utilize deep learning methodologies and tools for drug repurposing. We first summarized the commonly used bioinformatics and pharmacogenomics databases for drug repurposing. Next, we discuss recently developed sequence-based and graph-based representation approaches as well as state-of-the-art deep learning-based methods. Finally, we present applications of drug repurposing to fight the COVID-19 pandemic, and outline its future challenges.

## Graphical/Visual Abstract and Caption

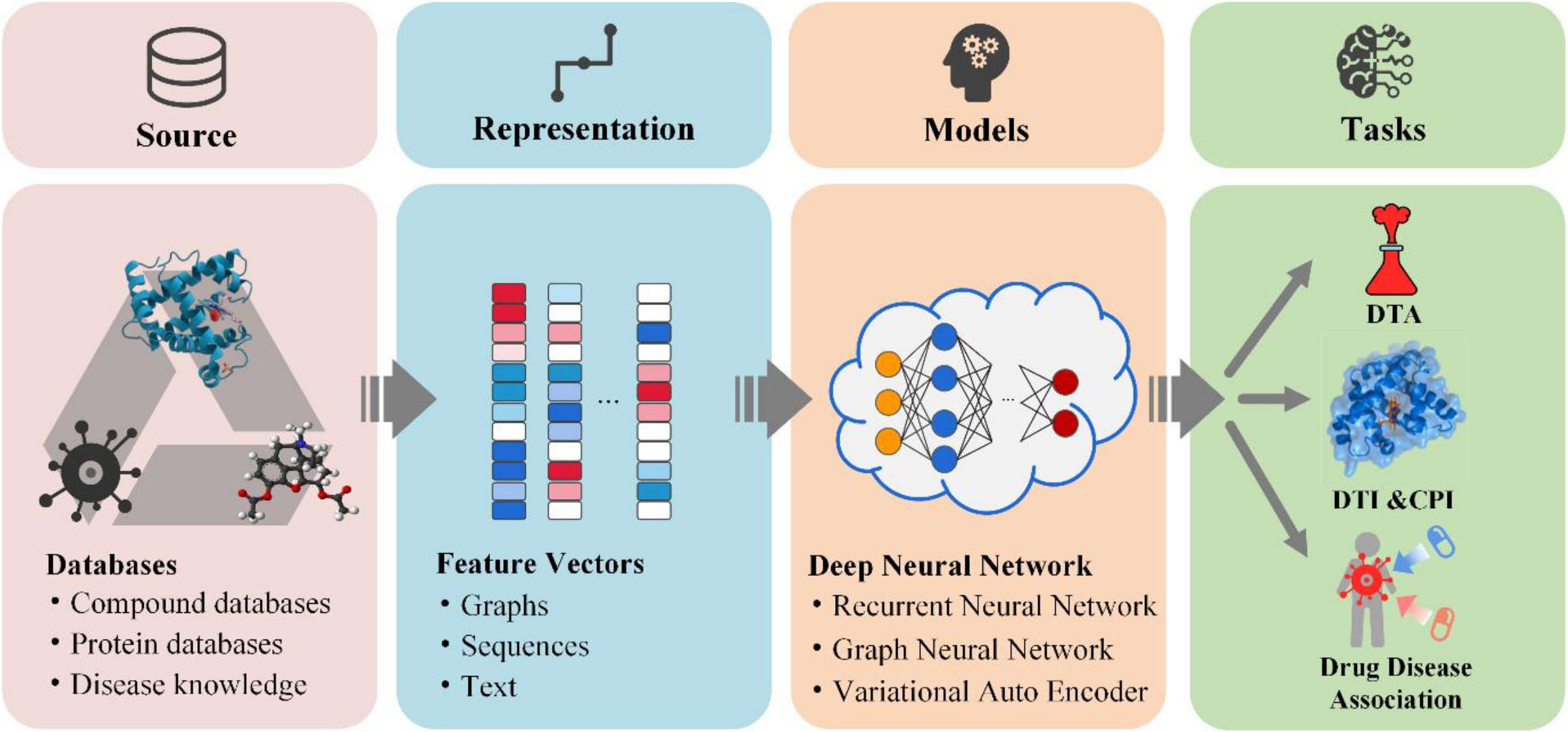

**FIGURE 1** A diagram illustrating the workflow of drug repurposing using deep learning approaches. The entire deep learning-based drug repurposing pipeline includes four steps: (1) Create high-quality data sources among compounds, proteins and diseases; (2) Generate informative feature vectors using various representation approaches (such as graphs, sequences, and text); (3) Build and evaluate various deep learning models; and (4) Conduct drug repurposing tasks, including prediction of drug-target binding affinity of (DTA), drug-target interaction (DTI), compound-protein interaction (CPI), and drug-disease associations.

## 1. INTRODUCTION

The development of new drugs is time-consuming and costly. According to the Eastern Research Group (ERG),[1] it usually takes 10-15 years and 0.8-1.5 billion dollars to develop a candidate drug, while the success rate of developing a new molecular entity is only 2.0%.[2] Effective identification of new indications from approved or well-established clinical drugs plays an essential part in drug discovery.[3]-[6] Such a process is also known as drug repositioning or reprofiling, and it can bypass many preapproval tests required for newly developed therapeutic compounds.[7] In general, drug repurposing offers a variety of advantages during the process of drug discovery, such as lower risk of failure, less investment and shorter development time frame.[8],[9]

In drug repurposing pipelines, machine learning methods take advantage of manually constructed descriptors to better predict the downstream tasks (e.g., molecular properties), which could identify possible candidates for subsequent clinical trials.[10] However, since these methods can only handle fixed-size inputs, early on most machine learning methods heavily depended on feature engineering[11] and domain knowledge. More recently, there has been a steady increase in the amount of available chemical and biomedical data in drug discovery and development. How to effectively explore the large-scale domain data becomes a critical task in drug repurposing. On the other hand, deep learning[12] has achieved remarkable success in a wide range of complex tasks, including natural language processing,[13] speech recognition[14] and computer vision.[15] Recently, deep learning methods have also started to be applied to drug repurposing.[16],[17] Different from traditional machine learning techniques, the strength of deep learning lies in its ability to learn complex relationships between input features and output decisions from large-scale data, helped by the continuous increase of computing power from hardware equipment. In particular, deep learning methods can automatically learn multiple levels of representations exclusively from its input data, without the need of additional user input. Although their applications in drug repurposing are still in the infancy stage, they have already shown great potential (**Figure 1**).

The past few years have seen a surge in drug repurposing research due to the unprecedented success of deep learning. Numerous methods, databases and applications have been proposed in the literature, calling for a comprehensive survey to focus the efforts in this flourishing new direction. While there have been several recent reviews focusing on computational methods,[9] they also cover machine learning and artificial intelligence (AI) algorithms,[18] including network-based approaches,[19] as well as recently developed deep learning models.[20] Some surveys of databases and other resources supporting drug repurposing have also been conducted recently.[21],[22] Other comments and opinions emphasize the drug design,[23] development process,[24] and the calculation methods in cancer research[25] or COVID-19 drug repurposing.[3] To our knowledge, no review has yet summarized and integrated these methods from a general point of view. Therefore, this review fills the gap by surveying drug repurposing approaches with a focus on recent developments in representation methods and deep learning models. We first summarize the widely used databases related to drug repurposing. Then we provide a brief overview of sequence-based and graph-based representation methods, respectively. Moreover, we investigate two kinds of drug repurposing deep learning models, target-based and disease-based. We also provide a comprehensive overview of several applications of drug repurposing techniques, including Coronavirus Disease 2019 (COVID-19), an infectious disease caused by severe acute respiratory syndrome coronavirus 2 (SARS-CoV-2). Finally, we highlight the challenges facing future developments of deep learning in drug repurposing.

## 2. DATABASE

The explosive growth of large-scale genomic, phenotypic, and omics data, provides computational drug repurposing approaches vast opportunities for the discovery of new candidate drugs.[26] Available databases also include potential cellular targets for different families of chemical compounds. For example, KEGG (Kyoto Encyclopedia of Genes and Genomes)[27] is an integrated database that contains large-scale molecular data sets from genes, proteins, biological pathways and human diseases, which is used for better understanding of high-level functions and applications of the biological system. DrugBank[28] is a comprehensive database that combines detailed drug information with the corresponding drug targets. The latest released version, DrugBank (V5.1.7) consists of 13,791 drug entries, which contain 2,653 small molecule drugs approved by Food and Drug Administration (FDA). Pubchem[29] is a database of chemical molecules and their activities against biological assays. It consists of three dynamically growing primary databases, including 110 million compounds, 271 million substances, and 297 million bioactivities so far. Here, we provide a brief summary of common databases involved in drug repurposing in **Table 1**.

**Tables 1** The widely-used databases in drug repurposing.

| Database | Describe | URL | Ref | API |
|---|---|---|---|---|
| BindingDB | A public database of protein-ligand binding affinities. | http://www.bindingdb.org/bind | [30] | √ |
| CCLE | Cancer Cell Line Encyclopedia (CCLE) is a large cancer cell line collection that broadly captures the genomic diversity of human cancers and provides valuable insight into anti-cancer drug responses. | https://portals.broadinstitute.org/ccle | [31] | |
| CellMinerCDB | An interactive web application that simplifies the access and exploration of cancer cell line pharmacogenomic data across different sources. | https://discover.nci.nih.gov/cellminercdb/ | [32] | |
| ChEMBL | A manually curated database of bioactive molecules with drug-like properties. It brings together chemical, bioactivity and genomic data to aid the translation of genomic information into effective new drugs. | https://www.ebi.ac.uk/chembl/ | [33] | √ |
| ChemDB | It provides chemical structures and molecular properties. ChemDB also predicts 3D structures of molecules. | http://cdb.ics.uci.edu/ | [34] | |
| ChemicalChecker | It provides processed, harmonized and integrated bioactivity data. | https://chemicalchecker.org/ | [35] | √ |
| CGI | Cancer Genome Interpreter (CGI) supports the identification of tumor alterations that drive the disease and flag those that may be therapeutically actionable. | https://www.cancergenomeinterpreter.org/ | [36] | |
| CTD (Comparative Toxicogenomics Database) | Comparative Toxicogenomics Database (CTD) provides manually curated information about chemical-gene or protein interactions, chemical-disease and gene-disease relationships. | http://ctdbase.org/ | [37] | |
| DGIdb | Drug–target interactions mined from > 30 trusted sources, including DrugBank, PharmGKB, Chembl, Drug Target Commons, Therapeutic Target Database. | http://www.dgidb.org/ | [38] | √ |
| DisGeNET | It is a discovery platform containing publicly available collections of genes and variants associated with human diseases. | http://www. disgenet.org/ | [39] | √ |
| DrugBank | It combines drug data (i.e., chemical, pharmacological and pharmaceutical) information with drug target information (i.e., sequence, structure and pathway). | http://www.drugbank.ca | [28] | √ |
| DrugCentral | It provides information on active chemical entities and drug mode of action. | http://drugcentral.org/ | [40] | √ |
| DTC | Drug Target Commons (DTC) manually curates bioactivity data along with protein classification into superfamilies, clinical phase and adverse effects as well as disease indications. | http://drugtargetcommons.fimm.fi/ | [41] | √ |
| DTP | Drug Target Profiler (DTP) contains drug target bioactivity data and implements network visualizations. DTP also contains cell-based response profiles of the drugs and their clinical phase information. | http://drugtargetprofiler.fimm.fi/ | [42] | |
| GeneCards | Automatically integrates gene-centric data from 150 web sources, including genomic, transcriptomic, proteomic, genetic, clinical and functional information. | https://www.genecards.org/ | [43] | |
| GLIDA | It contains drug-target interactions for G-protein-coupled receptors (GPCRs). | http://pharminfo.pharm.kyoto-u.ac.jp/services/glida/ | [44] | |
| GtopDB | It contains quantitative bioactivity data for approved drugs and investigational compounds. | http://www.guidetopharmacology.org/ | [45] | √ |
| KEGG | It is a knowledge base for systematic analysis of gene functions, linking genomic information with higher order functional information. | http://www.genome.jp/kegg | [27] | √ |
| LINCS | It contains details about the drug assays, cell types, and perturbagens that are currently part of the library, as well as software that can be used for analyzing the data. | http://www.lincsproject.org/LINCS/ | [46] | √ |
| OMIM | It is a comprehensive, authoritative compendium of human genes and genetic phenotypes that is freely available and updated daily. The full-text, referenced overviews in OMIM contain information on all known mendelian disorders and over 16,000 genes, and it focuses on the relationship between phenotype and genotype. | https://www.omim.org/ | [47] | √ |
| PathBank | PathBank is designed specifically to support pathway elucidation and discovery in transcriptomics, proteomics, metabolomics and systems biology. | https://pathbank.org/ | [48] | |
| PathwayCommon | Pathways including biochemical reactions, complex assembly and physical interactions involving proteins, DNA, RNA, small molecules and complexes. | http://www.pathwaycommons.org/ | [49] | √ |
| PDSP Ki | It contains bioactivity data in terms of $k_i$ especially for GPCRs, ion channels, transporters and enzymes. | https://pdspdb.unc.edu/pdspWeb/ | [50] | √ |
| PharmGKB | It contains comprehensive data on genetic variation on drug response for clinicians and researchers. | https://www.pharmgkb.org/ | [51] | √ |
| Probes & Drugs Portal | A public resource joining together focused libraries of bioactive compounds (e.g., probes, drugs, specific inhibitor sets). | https://www.probesdrugs.org/home/ | [52] | |
| Pubchem | It provides varieties of molecular information including chemical structure and physical properties, biological | https://pubchem.ncbi.nlm.nih.gov/ | [29] | √ |
| | activities, safety and toxicity information, patents, literature citations and so on. | | | |
| STITCH | It stores known and predicted interactions of chemicals and proteins, and currently covers 9,643,763 proteins from 2,031 organisms. | http://stitch.embl.de/ | [53] | √ |
| Supertarget | A data resource is used for analyzing drug-target interactions and drug side effects. | http://bioinf-apache.charite.de/supertarget/ | [54] | |
| SwissTarget-Prediction | It contains information on predicted targets of drugs based on the similarity principle through reverse screening. | http://www.swisstargetprediction.ch/ | [55] | |
| TTD | Therapeutic Target Database (TTD) provides information about the known and explored therapeutic protein and nucleic acid targets, the targeted disease, pathway information and the corresponding drugs directed at each of these targets. | https://db.idrblab.org/ttd/ | [56] | |

As shown in **Table 1**, the databases covered in this review can be divided into four main categories, including chemical, biomolecular, drug-target interaction and disease databases. To better utilize these data, the primary consideration is to focus on datasets that are publicly available online, where the associated data is either easy to download or easy to get access to via an API (Application Programming Interface). This criterion is crucial as it allows data to be easily integrated into a deep learning method. Then researchers should select the desired input from various data sources or cross database comparative analysis. For example, DrugBank provides the drug-target interaction data, which can be obtained by reading its description and checking the data statistics, but it also provides clinical, drug classification, chemical structures, pathways and drug combination information. The more details of databases providing drug repositioning information can refer to Tanoli et al.'s study.[22] Actually, deep learning is well suited to integrate heterogeneous data sources. A recent study named CDRscan[57] was proposed to integrate with the genomic data from CCLE, drug response assay data from GDSC, virtual docking based on structural fingerprints, and quantitative structure-activity relationships (QSAR) information from DrugBank. It can predict the anti-cancer activity from 1,487 approved drugs, which results in 14 oncology and 23 non-oncology drugs having new potential cancer indications. Additionally, considering more omics data can further create new opportunities for in silico drug repositioning. AOPEDF[58] was designed to collect physical drug-target interactions from DrugBank, TTD and PharmGKB, respectively, and to leverage the bioactivity data for drug-target pairs from ChEMBL and BindingDB, and to extract the chemical structure of each drug with SMILES format from DrugBank. It cleaned up the data according to the unique UniProt accession number and the threshold of binding affinity, to construct a heterogeneous network covering chemical, genomic and phenotypic data sources. A cascade deep forest classifier was built to infer new DTIs, which achieved high accuracy on two external validation sets collected from DrugCentral and ChEMBL. In comparison to the methods that rely on constructing the complex features by using matrix factorization or network modeling; herein, we only concentrate on deep learning methods that depend on the raw data (i.e., SMILES for drug and protein sequence for target), which can automatically extract the molecular features by designing the efficient representation learning (i.e., sequence-based and graph-based methods).

## 3. REPRESENTATION LEARNING

Inspired by the great success of deep learning in many scientific fields, including life science,[59] researchers have become increasingly interested in applying deep learning methods to computational drug repurposing, thereby saving time and cost. As a branch of machine learning, deep learning combines artificial neural networks with multiple layers of non-linear processing units to progressively extract high-level features from the raw input.[12] In fact, the performance of deep learning methods is largely reflected in the effective data representation, which means that a system can be allowed to automatically discover the representations required for feature extraction or classification from raw data using a set of techniques. Such a process is known as representation learning, and it is one of the fundamental steps in end-to-end deep learning.[60] Therefore, many efforts have been made to integrate deep learning methods into the design of feature representations of the input data that make it easier to extract useful information.[61] Representation learning used in drug repurposing can be mainly categorized into sequence-based and graph-based methods, respectively.

### 3.1 Sequence-based representation

Sequence-based representation methods can partly overcome the limitations of available protein/target structural data and the requirements of costly molecular docking simulation, while available biological data of protein and compound sequences offer a possibility for the rapid advancement of drug repurposing. For molecular compounds, one critical one-dimensional (1D) representation is SMILES (Simplified Molecular Input Line Entry System),[62] a text notation for the topological information based on chemical bonding rules (**Figure 2a**). In addition, chemical fingerprints, such as circular fingerprints,[63] are a 2D representation of molecules, which recurrently search for the partial structures around each atom, and then use a hash function to convert the molecule into a binary vector (**Figure 2b**). However, since the generated vectors are not only high-dimensional and sparse, they might contain 'bit collisions' owing to the hashing function. Recently, representation learning brought several breakthroughs in compound space. Specifically, Recurrent Neural Network (RNN) and Convolutional Neural Network (CNN) models are adopted to automatically learn latent features from SMILES strings to achieve better performance.[16],[17] Inspired by the pre-trained language model in Natural Language Processing (NLP),[64] Mol2vec[65] was proposed and recognized as the most representative method that considers molecular substructures as "words" and compounds as "sentences", and generates the embedding of atom identifiers by using Word2Vec.[66] Although these methods achieve excellent performance, the obvious disadvantage of such 1D or 2D representation is that information about bond lengths and 3D conformation is lost, which may be important for the binding detail of drug target. Therefore, the 3D representation will attract more attention in the future.

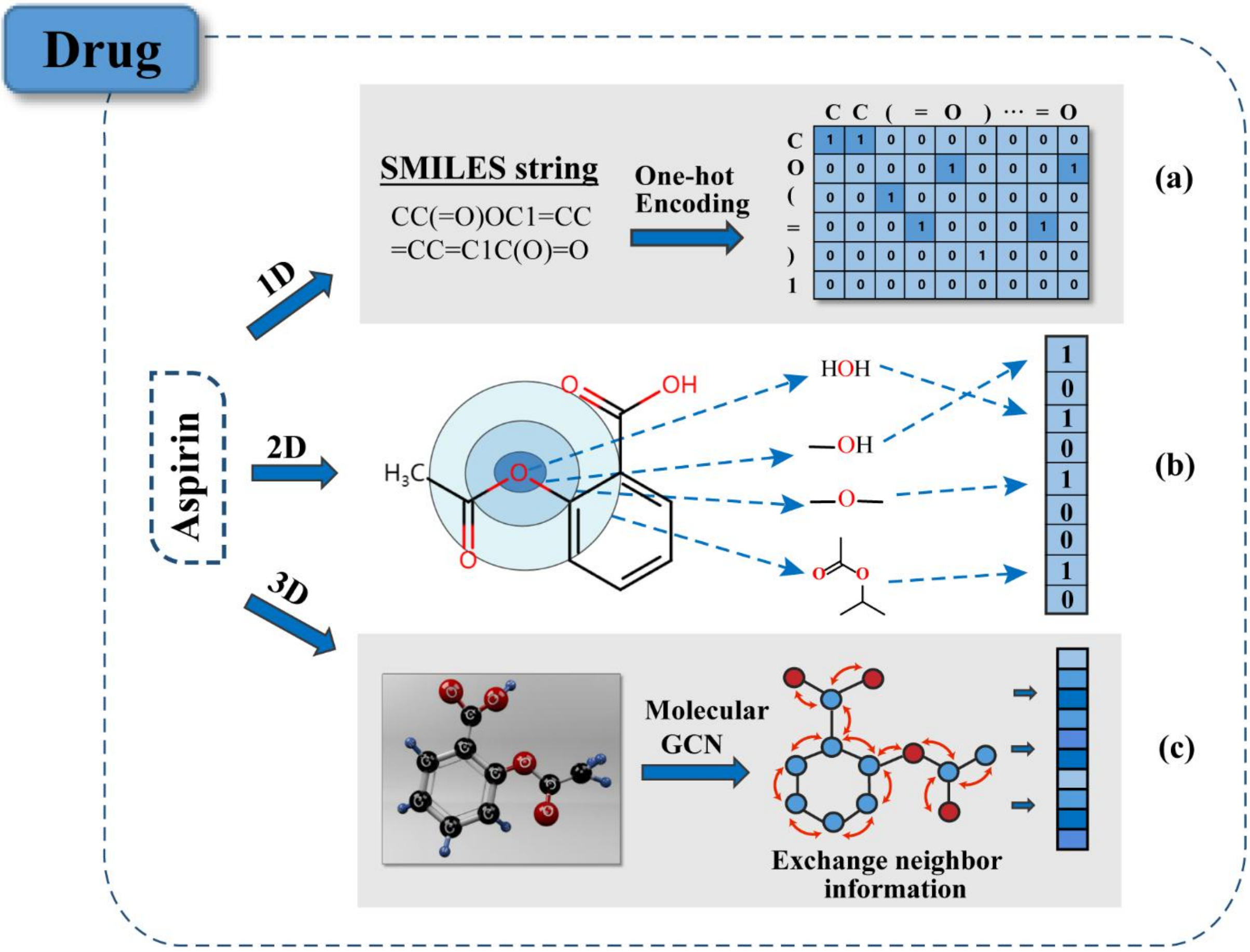

**FIGURE 2** Drug representations. (a) One-hot representation[67] of SMILES string. (b) Two-dimensional (2D) representation of molecular graph where each substructure was associated with a predefined bit vector. (c) Graph Neural network (GNN) was adopted to transfer a molecular graph to a vector where the atoms and bonds were denoted by nodes and edges, respectively.

Similarly, protein sequences are generally composed of 20 standard amino acids, where each amino acid can be simply encoded by one-hot encoding (**Figure 3a**). Besides, proteins can also be represented with a two-dimensional (2D) distance map (**Figure 3b**), which calculates the distance between all possible amino acid residue pairs in a three-dimensional protein structure.[68] Inspired by the embedding techniques of NLP, ProtVec[69] and doc2vec[70] were further developed to generate the non-overlapping 3-gram sub-sequences from protein sequences, and to pre-train their distributed representations based on a skip-gram model by using the word2vec technique. However, these models usually focused on learning context-independent representations. Different from k-gram, a unified representation method[71] was designed to apply RNN to learn statistical representations of proteins from unlabeled amino acid sequences, which are semantically rich and structurally, evolutionarily and biophysically grounded. Strodthoff et al.[72] proposed a universal deep sequence model which was pre-trained on unlabeled protein sequences and could be fine-tuned on downstream classification tasks. However, the protein representations mentioned above use only the information provided by the special order of the protein sequence consisting of 20 different characters, ignoring the physical, chemical and biological properties of the protein. Rifaioglu et al. proposed a new featurization method to represent protein sequences as digital matrices, based on their physical, chemical and biological properties.[73] Similar to compounds, the sequence-based representation methods do not take into account more information about the three-dimensional structure of the protein. Last but not least, a deep learning system named AlphaFold,[74] developed by Google DeepMind, has released the predicted 3D structure of a protein-based solely on its genetic sequence, which can take months by traditional experimental approaches. More recently, DeepMind further released the source code of AlphaFold2[75], and then their predicted 3D structures of human proteins are freely available to the community via a public database.[76]

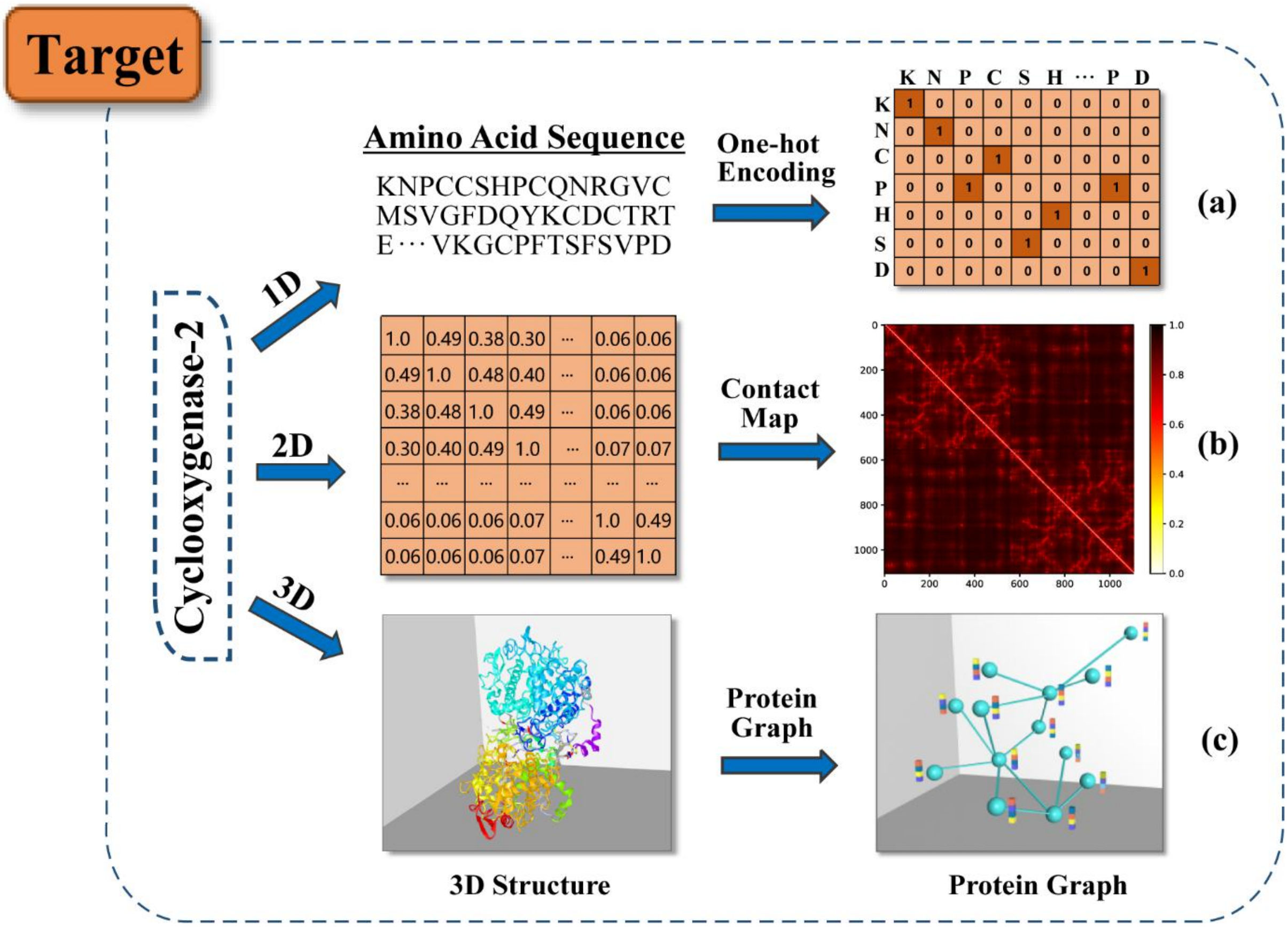

**FIGURE 3** Target representations. (a) One-hot representation of amino acids sequences. (b) Contact map was a kind of two-dimensional (2D) representation of the protein. (c) Graph convolutional network was used to learn the representation of the three-dimensional (3D) protein graph with nodes representing the various constituent non-hydrogen atoms.

### 3.2 Network/graph-based representation learning

Recent advancements of multi-omics technologies and systems biology approaches have generated large-scale heterogeneous biological networks, which provide considerable opportunities for graph or network-based drug repurposing.[77] Owing to the topological structure of the graph itself, and since a compound or a protein can be naturally encoded as a graph or a network, including their chemical associations, graph-based representation approaches have increasingly become an emerging solution to improve the performance in drug repurposing.

More recently, graph neural network (GNN)[78] has been developed as the state-of-the-art method for graph-related tasks, such as node-level and graph-level classification.[79],[80] Its advantage is in automatically extracting the latent features by considering the structure of neighboring nodes and aggregating the information among layers. SMILES string can be easily transformed into a molecular graph by RDKit.[81] For molecules, we can represent the atoms and bonds as vertices connected by edges[82] **(Figure 2c)**. For proteins, a more natural way to represent a protein molecule is to encode a protein graph with nodes representing the various constituent non-hydrogen atoms in the protein, a representation whose construction is rotationally invariant. ProteinGCN[83] effectively utilized both inter-atomic orientations and distances, and also captured the local structural information through the graph convolution formulation (**Figure 3c**). Compared to those GNNs that mainly retain first-order or second-order proximity, another promising technique, named network embedding, is used to learn the global features. Specifically, it usually maps nodes, edges, and their features to a vector, which maximally preserves global properties (*e.g.*, structural information).[84] Once the node representation is obtained, deep learning models can be applied to network-based tasks, including node classification,[85] node clustering[86] and link prediction.[87] Another important graph-based deep learning method, called the probabilistic graph, combines a variety of neural generative models, gradient-based optimization and neural inference techniques. Furthermore, variational autoencoders (VAE)[88] trained on biological sequences have been shown to learn biologically meaningful representations beneficial for various downstream tasks. In short, VAE is the variant of autoencoder that provides a stochastic map between the input space and the latent space. This map is regularized during the training to make sure that its latent space has the ability to generate some new data. An example of applying VAE in the protein modeling field is learning a representation of bacterial luciferase.[89] The resulting continuous real-valued representation can then be used to generate novel, functional variants of the *luxA* bacterial luciferase.

## 4. DEEP LEARNING MODELS FOR DRUG REPURPOSING

A drug repurposing tool usually aims at predicting unknown drug-target or drug-disease interactions, which can be either classified as “target-centered” or “disease-centered” methods, respectively. The target fishing strategy[90] encoded the chemical structure of drugs to screen targeted proteins, which provide the detailed poly-pharmacological interpretation. However, a single predicted target cannot fully describe the characteristics of the disease. Thus, effectively identifying the associations between drugs and diseases becomes essential for understanding the underlying biological mechanisms. Each approach presents the unique challenges of informatics, and this review focuses on the target-based and disease-based deep learning methods for drug repurposing over the last few years, respectively. **Table 2** Summary of details for all selected methods.

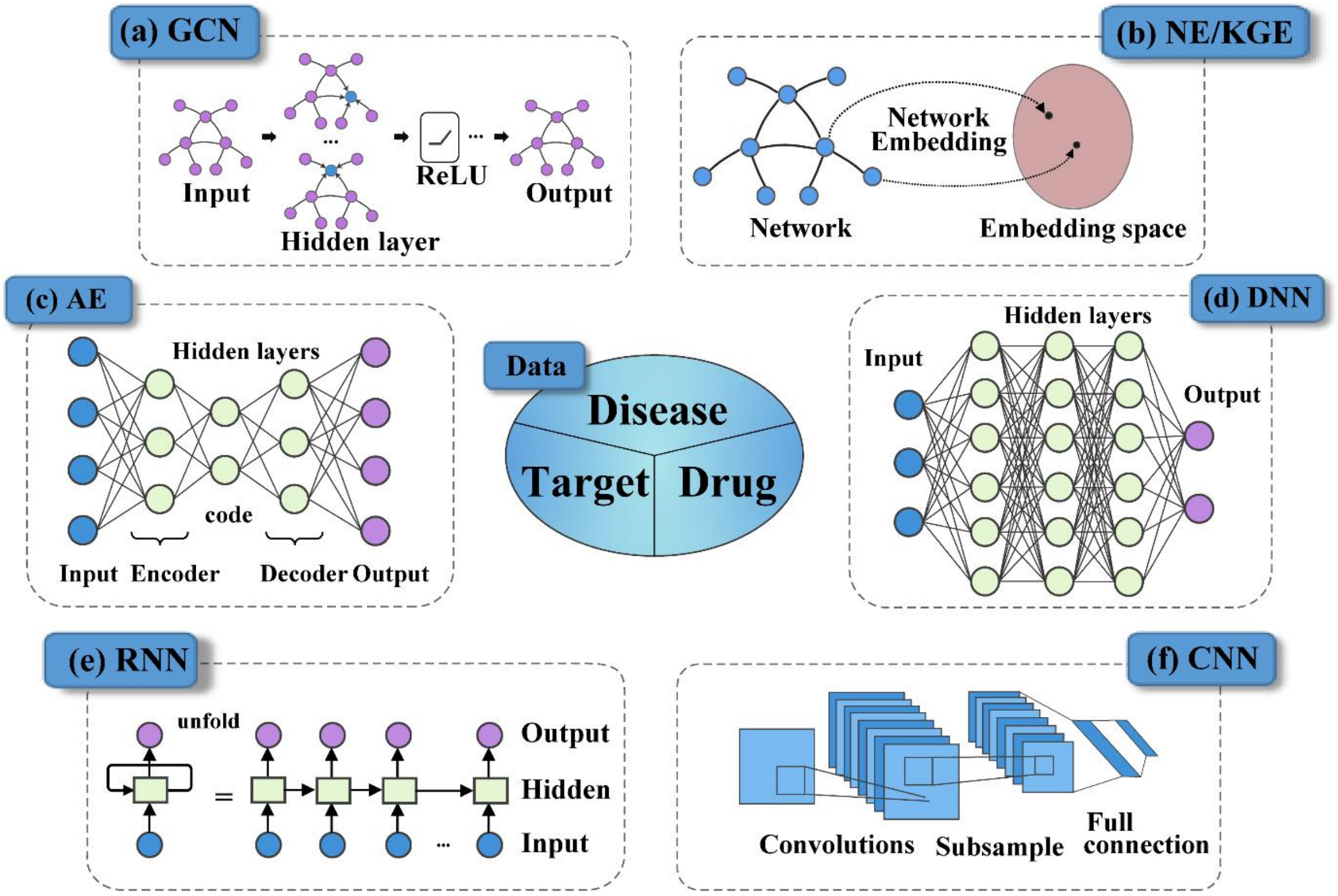

**FIGURE 4** Architecture of deep-learning models. (a) Graph convolutional network (GCN). (b) Network embedding (NE) or knowledge graph embedding model (KGE).[91] (c) Autoencoder (AE). (d) Fully connected deep neural network (DNN). (e) Recurrent neural network (RNN). (f) Convolutional neural network (CNN).

### 4.1 Target-centered models

Many deep learning methods have been exploited to find potential drug-target interactions based on molecular structure. Specifically, convolutional operations were used to perform on various lengths of amino acids sequences, and to capture the local residue patterns of generalized protein classes that play a critical part in drug-target interaction (DTI) prediction.[92] In order to make full use of compound-protein interaction (CPI) data, a novel multi-channel PCM-based DNN (**Figure 4d**) framework named MCPINN was proposed to predict DTIs.[93] In particular, it utilized three modules including feature extractor, end-to-end learner and classifier. It took compound SMILES, ECFPs and vectors embedded by Mol2vec,[65] as well as amino acid sequences embedded by ProtVec[69] as the input.

Recently, there have been numerous methods encoding compounds as molecular graphs. Tsubaki et al.[94] developed a novel end-to-end approach for CPIs prediction, by combining a GNN for compounds and a CNN (**Figure 4f**) for proteins, which could learn low-dimensional real-valued vector representations of molecular graphs and protein sequences. Similarly, Gao et al.[95] utilized LSTMs and GCN (**Figure 4a**) to project proteins and drug structures into dense vector spaces, respectively, and a two-way attention mechanism[96] was used to calculate how the pairs interact and thus enable interpretability. However, sequence-based CPIs models still have several limitations, such as splitting methods as well as hidden ligand bias, which results in overestimated prediction performance.[98] To address these limitations, a transformer architecture[97] with a self-attention mechanism named TransformerCPI was proposed, in which GCNs were employed to learn the representation of each atom, and proteins are transformed into sequential representations by embedding.[98] Compared with previous models, TransformerCPI achieved the best performance in more rigorous label inversion experiments.

Network-based approaches have been adopted for target identification for known drugs to assist in countering side effects and accelerate drug repurposing. For example, Luo et al.[99] first followed an unsupervised approach to learn low-dimensional vector representations of drugs and targets from heterogeneous networks, and then adopted inductive matrix completion to predict novel DTIs. However, separating feature learning from a prediction task may not produce an optimal solution. Subsequently, the same group further proposed a neural network-based DTI prediction method, termed NeoDTI.[100] NeoDTI integrated neighborhood information of nodes in the heterogeneous network and automatically learned topology-preserving representations of drugs and targets. However, these methods are prone to preserving only the local proximity. Thus, deep autoencoder was adopted to automatically learn high-quality features from heterogeneous networks, and Zeng et al.[77] employed positive-unmarked matrix completion to predict new DTIs, named deepDTnet, which integrates large biomedical network datasets for target identification and minimizes the translational gap in drug development. The comparative experiments show that the proposed deepDTnet achieves a high AUC-ROC metric of 0.963 in identifying novel molecular targets for known drugs, outperforming traditional machine learning approaches, including random forest (0.911), SVM (0.869), k-nearest neighbors (0.839), and Naive Bayes (0.783). Further, different from deepDTnet, a deep learning-based framework, named AOPEDF, an arbitrary-order proximity embedded deep forest,[58] was proposed to predict the DTIs. Specifically, it constructed 9 networks for drugs, including clinically reported drug-drug interactions, to consider the complementary order proximity information for different networks, and it also achieved higher performance with fewer hyperparameters. Additionally, it showcased in a case study that multiple molecular targets predicted by AOPEDF are related to mechanism-of-action of substance abuse disorder for several marketed drugs (e.g., aripiprazole, risperidone and haloperidol), which were successfully supported by experimental assays. It further analyzed the potential advantages through ablation experiments. Especially, it was replaced by LINE[101] (i.e., $LINE_{1st}$ and $LINE_{2nd}$) for feature extraction, and the designed deep forest classifier was compared with the traditional methods equipped with the same features, including Support Vector Machine, Random Forrest and Deep Neural Network, where the results showed that the high-order proximities preserved by AROPE may provide more effective information for classification, and deep forest classifier achieved the best performance. The currently available knowledge bases were used to generate a knowledge graph (KG) of biological entities, and a specific KG embedding model called Trimodel was employed to learn low-dimensional vector representations of drugs and targets, respectively. Naturally, the DTIs prediction can also be modeled as link prediction in KG.[87]

Most of the studies mentioned above have focused on binary classification, where the goal was to determine whether a drug-target pair interacts or not. While the protein-ligand interactions (PLIs) can predict a binding affinity value, it's more challenging for drug repurposing. For example, with the novel representations of structurally annotated protein sequences (SPS), Karimi et al. proposed a semi-supervised deep learning model called DeepAffinity.[16] DeepAffinity unified RNN (**Figure 4e**) and CNN to jointly encode molecular representations, and predicted affinity using both unlabeled and labeled data. What's more, DeepAffinity introduced some attention mechanisms[96] to interpret predictions by separating molecular fragments or their major contributors, which can be further applied to predict the binding sites and sources of binding specificity. GraphDTA[17] was also applied to predict DTA (drug-target binding affinity), but the difference was that GNN was used instead of CNN to learn the representation of the compound. However, in the above methods, the physical, chemical and biological properties of proteins are generally ignored. Therefore, Rifaioglu et al.[73] proposed a novel featurization approach for proteins, which integrated multiple types of protein characteristics, such as sequence, structure, evolution, and physico-chemical properties, into a two-dimensional vector, and achieved significant improvements in terms of CPA (compound protein affinity) predictive performance.

### 4.2 Disease-centered models

Identifying the interactions between drug-disease pairs becomes essential for disease-centered drug repurposing. Currently, existing methods can be roughly categorized into similarity-based and network-based approaches.

Previous methods have been proposed to calculate the similarity between drugs and diseases. The methods have achieved certain success in computational drug repurposing, by combining drug or disease features with the known drug-disease associations. For example, a robust approach termed SNF-CVAE[102] was developed to predict novel drug-disease interactions. Specifically, it integrated similarity measurement, similarity selection, similarity network fusion (SNF) and collective variational automatic encoder (CVAE)[103] for nonlinear analysis, which improved the accuracy of drug-disease interaction prediction. Meanwhile, it showcased in two case studies that the top drug candidates predicted by SNF-CVAE can potentially treat Alzheimer's disease and Juvenile rheumatoid arthritis, which were successfully validated by clinical trials and published studies. Furthermore, Xuan et al.[104] proposed a novel method based on CNN and bidirectional LSTM for drug repurposing, where the CNN-based module was used to learn the original representation of drug-disease pairs from their similarities and associations; yet, the BiLSTM-based module was used to learn the path representations of the drug-disease to balance the contributions of different paths by attention mechanism.

**TABLE 2** Drug repurposing methods based on deep learning.

| **Target-centered models** | | | | | |
|---|---|---|---|---|---|
| Model | input | | network architecture | type | year |
| | protein | compound | | | |
| DeepAffintiy[16] | Protein SPS (Structural property sequence) | SMILES | RNN, CNN, Attention Mechanism | DTA | 2019 |
| Rifaioglu et al.[73] | Protein sequence, structural, evolutionary and physicochemical properties | SMILES | CNN | DTA | 2020 |
| GraphDTA[17] | Protein sequence | Molecular graph | GCN, CNN | DTA | 2019 |
| DeepConv-DTI[92] | Protein sequence | Fingerprint | CNN, DNN | DTI | 2019 |
| MCPINN[93] | Amino acid sequence & ProtVec | ECFP & Mol2Vec & SMILES | DNN | CPI | 2019 |
| Gao et al.[95] | Amino acid sequence | Molecular graph | GCN, LSTM, two-way Attention Mechanism | DTI | 2018 |
| TransformerCPI[98] | Protein sequence | Molecular graph | Transformer | CPI | 2020 |
| Tsubaki et al.[94] | Amino acid sequence | Molecular graph | GCN, CNN, Attention Mechanism | CPI | 2019 |
| NeoDTI[100] | Eight individual drug or target-related networks | | GCN | DTI | 2019 |
| DeepDTnet[77] | 15 types of chemical, genomic, phenotypic, and cellular networks | | Autoencoder | DTI | 2020 |
| AOPEDF[58] | 15 networks covering chemical, genomic, phenotypic and network profiles among drugs, proteins/targets and diseases. | | Deep forest algorithm | DTI | 2020 |
| Trimodel[87] | Biomedical knowledge graphs about drug and target | | Knowledge Graph Embedding | DTI | 2019 |
| **Disease-centered models** | | | | | |
| | input | | network architecture | type | year |
| Model | Drug & Disease | | | | |
| SNF-CVAE[102] | Drug-related similarity information, Drug-Disease associations | | Similarity Network Fusion (SNF), Collective Variational Autoencoder (cVAE) | Drug-disease association prediction | 2020 |
| Xuan et al.[104] | Drug network, disease network, Drug-Disease associations | | CNN, BiLSTM | Drug-disease association prediction | 2019 |
| DeepDR[105] | Drug-Disease, Drug-Side-effect, Drug-Target and seven Drug-Drug networks | | Multimodal deep autoencoder (MDA), Collective Variational Autoencoder (cVAE) | Drug-disease association prediction | 2019 |
| Wang et al.[106] | Drug-Protein, Disease-Protein and PPIs | | Bipartite GCN | Drug-disease association prediction | 2020 |
| Cov-KGE[108] | Biomedical knowledge graphs | | Knowledge graph embedding | Drug-disease association prediction | 2020 |

On the other hand, network-based approaches represent graph information among different biological networks to boost the performance of drug repurposing. For example, Su et al.[84] summarized the use of network embedding (**Figure 4b**) methods in biomedical data and discussed a broad range of potential applications and limitations. Furthermore, a network-based deep-learning method, termed deepDR,[105] was developed for *in silico* drug repurposing. Specifically, it firstly learned high-level features of drugs from 10 networks via a multi-modal deep autoencoder. Then combined with the clinically reported drug-disease pairs, the learned drug representations were encoded and finally decoded by a variational autoencoder (**Figure 4c**) to infer the candidates for approved drugs. Compared with conventional network-based and machine learning-based approaches, including DTINet, KBMF, Random Forest and Support Vector Machine, the proposed deepDR achieved an AUROC score of 0.908 with 4.6% absolute gain compared to DTINet (the second-best method). Importantly, it showcased that the top 20 candidates predicted by deepDR are approved for the treatment of Alzheimer's disease (e.g., risperidone and aripiprazole) and Parkinson's disease (e.g., methylphenidate and pergolide), most of them can be validated by previous literature. However, deepDR only considered information sources in the drug domain rather than the interactions in the disease domain. Wang et al.[106] assembled interactions across protein, drug and disease domains from large-scale databases, which provides insights into utilizing protein-protein interactions (PPIs) for improved drug repurposing assessment. Specifically, a bipartite GCN-based method was designed to merge with inter-domain information. Also, a biological system can be modeled by using heterogeneous multi-relational networks (i.e., knowledge graphs). In another study, Mohamed et al.[107] exclusively explored knowledge graph embedding (KGE) models, focused on performing the best models in terms of both scalability and accuracy across various biological tasks, and further discussed the opportunities and challenges of using KGE to model biological systems. In detail, Zeng et al. built a comprehensive knowledge graph that includes entities of drugs, diseases and proteins/genes from a large scientific corpus of 24 million PubMed publications. A powerful Cov-KEG model was used to quickly identify drugs that can be repurposed for the potential treatment of COVID-19.[108] However, to a certain extent, the potential noise from different data sources and the sparseness of the data will affect the performance of the knowledge graph method.

### 4.3 Model evaluation

Deep learning models are usually evaluated by cross-validation, which involves partitioning the original observation dataset into a training set for model training, and an independent set used to evaluate the model performance. K-fold cross-validation is the widely-used cross-validation technique. Meanwhile, drug repurposing tasks are roughly divided into two categories, including classification and regression. As for regression tasks, Root Mean Squared Error (RMSE), Mean Absolute Error (MSE) and the concordance index (CI) are adopted to access the model performance. Specifically, MSE represents the sum of the absolute differences between predictions and actual values.[17] RMSE measures the average magnitude of the error by taking the square root of the average of squared differences between prediction and actual observation.[73] On the other hand, CI measures the probability of two randomly selected compound-target protein pairs with different binding affinity values to be in the correct order.[73] As for classification tasks, accuracy, the area under the receiver operating characteristic curve (AUC-ROC), the area under the precision-recall curve (AUPR) and the F1-score are often used to evaluate the performance of the classifiers. Specifically, accuracy defines overall accuracy as the probability of correspondence between a positive decision and true condition. AUC-ROC is a metric for measuring the ability of a binary classifier to discriminate between positive and negative classes.[77] While especially for highly skewed data, AUC-ROC may be overly optimistic in evaluating the performance of prediction algorithms, AUPR can provide a better assessment in this case. A PR curve shows the trade-off between precision and recall across different decision thresholds.[77] F1-score is a measure of a test's accuracy. It is calculated from the precision and recall of the test, where the precision is the number of true positive results divided by the number of all positive results, including those not identified correctly, and the recall is the number of true positive results divided by the number of all samples that should have been identified as positive.[102] On the other hand, there are several systematic benchmarks and platforms to accelerate machine-learning model development, validation and transition into biomedical and drug discovery. For example, a large-scale benchmark for molecular machine learning, named MoleculeNet,[109] was developed to provide high-quality open-source implementations of multiple previously proposed molecular featurization and learning algorithms. Meanwhile, a comprehensive and easy-to-use deep learning library was designed to predict the drug-target interaction (DTI), it supports the training of customized DTI prediction models by implementing 15 compound and protein encoders and over 50 neural architectures, along with providing many other useful features.[110] More recently, the first unifying framework, named Therapeutics Data Commons (TDC),[111] was released to systematically access and evaluate machine learning across the entire range of therapeutics.

In addition to computational evaluation discussed here, experimental validations or clinical validation for a short list of high-confidence of predicted candidates. Commonly used experimental validation approaches include *in vitro* models and in vivo animal models. For example, a team experimentally validated that deepDTnet predicted topotecan (an approved topoisomerase inhibitor) is a new, direct inhibitor of human retinoic-acid-receptor-related orphan receptor-gamma t (ROR-γt). Subsequently, the same team showed that topotecan revealed a potential therapeutic effect in a mouse model of multiple sclerosis by specifically targeting ROR-γt. A classic clinical validation approach is case-control observational studies using electronic patient data generated from health insurance claims or electronic health records. Using retrospective case-control observations with 7.2 million individuals, a team identified that usage of fluticasone (an approved glucocorticoid receptor agonist) is significantly associated with a reduced incidence of Alzheimer's disease.[112] Using large healthcare databases with over 220 million patients and state-of-the-art pharmacoepidemiologic analyses, another team identified that hydroxychloroquine (an approved immunosuppressive drug) is associated with a decreased risk of coronary artery disease (CAD); furthermore, in vitro experiments show that hydroxychloroquine attenuates pro-inflammatory cytokine-mediated activation in human aortic endothelial cells, mechanistically supporting its potential beneficial effect.[5] In summary, combining computational prediction and experimental or clinical validation will offer actionable strategies to identify repurposable drug candidates to be tested in patients directly.

## 5. APPLICATIONS OF DRUG REPURPOSING

Due to many changes in traditional *de novo* drug discovery, drug repurposing has been demonstrated as a promising strategy for drug discovery and development in a variety of human diseases, such as rare diseases,[113]-[115] neurodegenerative disease,[112],[116]-[119] cancer,[120]-[122],[25] infectious disease.[3],[123],[124] In this Review, we will use COVID-19, an infectious disease caused by SARS-CoV-2, as an example, to highlight how drug repurposing strategies accelerate therapeutic development to fight the crisis of COVID-19 pandemic(**Figure 5**).

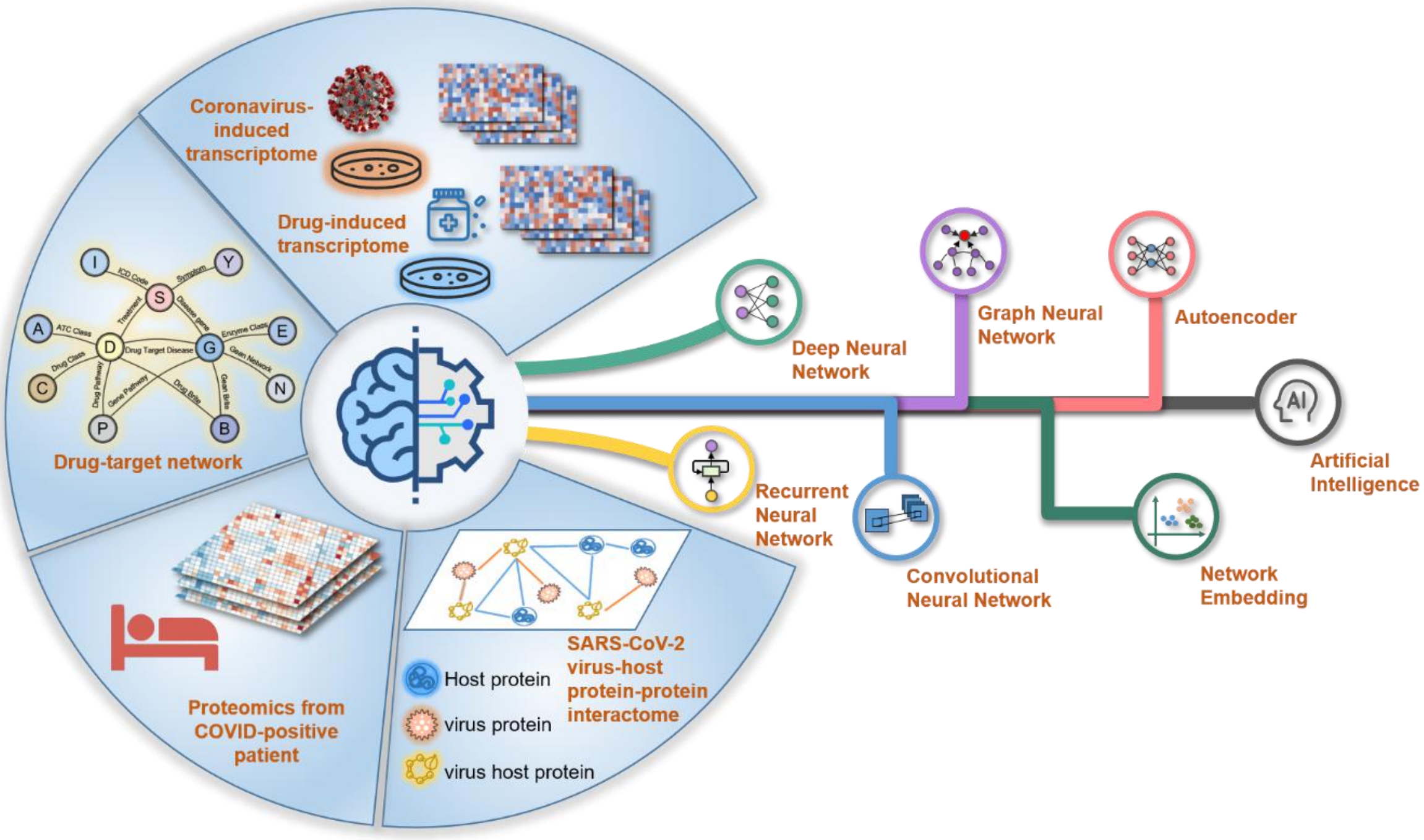

**Figure 5** A diagram illustrating deep learning-based drug repurposing infrastructure for emerging development of host-targeting therapies to fight COVID-19 and future pandemic. We posited that approved drugs that specific human proteins/targets may offer potential host-targeting therapies for COVID-19 as COVID-19 may share biology with human cells and tissues from the SARS-CoV-2 virus-host protein-protein interactome perspective.[3],[4],[6]

The ongoing global COVID-19 pandemic has led to more than 224 million confirmed cases and approximately 4 million deaths worldwide as of Sep 13, 2021. There are no proven effective therapies for COVID-19 although there are available vaccines. There is a critical, time-sensitive need to develop effective prevention and treatment strategies for the COVID-19 pandemic, including drug repurposing strategies.[3] For example, an autoencoder-based platform that systematically integrates available transcriptomic, proteomic and structural data was proposed.[125] The authors highlighted the importance of serine/threonine and tyrosine kinases as potential targets that intersect the SARS-CoV-2 and aging pathways and computationally prioritized several drug candidates (i.e., doxapram, dasatinib, and ribavirin) for old individuals with COVID-19.[125] In addition to host-targeting therapies for COVID-19, antiviral drug repurposing that specifically targets viral proteins of SARS-CoV-2 is also an attractive approach. For example, SARS-CoV-2 main protease (Mpro) is one of the most favorable drug targets. A study integrates mathematics (i.e., algebraic topology) and deep learning (termed MathDL) to provide a reliable ranking of the binding affinities of candidate inhibitors across 137 crystal structures of MPro.[126] The team computationally identified 71 candidate covalent bonding inhibitors of MPro of SARS-CoV-2 using MathDL.[126] Another team proposed a neural network-based method, termed DeepCE in which utilizes a graph neural network and multi-head attention mechanism[96] to predict chemical substructure-gene and gene-gene associations perturbed by compounds. The authors utilized a data augmentation method[127] that extracts useful information from unreliable experiments (i.e., Average Pearson Correlation (APC) score<0.7) in the L1000 dataset and showed high performances of DeepCE compared to several state-of-the-art methods.[128] The team further applied DeepCE to drug repurposing of COVID-19 and computationally prioritized a set of candidate compounds consistent with ongoing clinical evidence on COVID-19.[128]

Although these studies demonstrated the potential of deep learning approaches for possible identification of candidate repurposable drugs for COVID-19, including host-targeting therapies and antiviral treatments, some recent studies also achieve comparative or even better performance than deep learning methods on the application with SARS-CoV-2, by employing simpler strategies. For example, a web-based platform designed for SARS-CoV-2 virus-host interactome exploration and drug-target identification, termed CoVex,[129] was developed to implement systems medicine algorithms for network-based prediction of drug candidates, and to mine the integrated virus-host-drug interactome for putative drug targets and drug repurposing candidates. Meanwhile, a multimodal ensemble forecasting approach[130] was proposed to combine with artificial intelligence, network diffusion, and network proximity, and it experimentally screened in human cells the top-ranked drugs, which finally identifies four drugs (digoxin, fluvastatin, azelastine, and auranofin) that could be repurposed to potentially treat COVID-19. On the other hand, none of the predictions were validated by preclinical models and clinical randomized controlled clinical trials. It should therefore be noted that all predicted candidate drugs must be validated using experimental assays and randomized clinical trials before they can be recommended for use in patients with COVID-19.

A novel method named Cov-KGE[108] was proposed to develop an integrative and network-based deep learning methodology. Resulted from a large scientific corpus of 24 million PubMed publications, the CoV-KGE was adopted to build a comprehensive knowledge graph that includes 15 million edges across 39 types of relationships connecting drugs, diseases, proteins/genes, pathways, and expressions. Using the ongoing COVID-19 trial data as a validation set, they demonstrated that CoV-KGE had a high performance in identifying repurposable drugs for COVID-19, identified 41 high-confidence repurposable drugs (including dexamethasone[131] and melatonin) for COVID-19, which were validated by enrichment analysis of gene expression and proteomic data in SARS-CoV-2 infected human cells. Subsequently, the same team identified that melatonin usage is significantly associated with a 28% reduced likelihood of a positive laboratory test result for SARS-CoV-2 confirmed by reverse transcription-polymerase chain reaction assay, using a large COVID-19 registry database.[4] Currently, there are at least 8 ongoing or pending clinical trials to test the clinical effects of melatonin in the potential treatment of COVID-19 from the clinicaltrials.gov database (www.clinicaltrials.gov). Combining computational strategies (including deep learning) and real-world patient data validation will offer more promising candidate repurposable drugs to be tested in clinical trials shortly.[3] Using BenevolentAI's knowledge graph,[132] baricitinib was identified as a candidate agent for possible treatment of COVID-19. Several Phase II Randomized Double-Blind Trials of baricitinib or its combination therapy with available antiviral agents are under investigation for COVID-19 patients (ClinicalTrials.gov identifier: NCT04373044 and NCT04401579). Recently, baricitinib was associated with reduced mortality in hospitalized adults with COVID-19 in a phase 3, double-blind, randomised, placebo-controlled trial,[133] showing the first successful example of deep learning approaches for COVID-19 drug repurposing development.

As COVID-19 patients flood hospitals worldwide, physicians are trying to search for effective antiviral therapies to save lives. In summary, deep learning approaches offer promising strategies for the rapid development of effective therapeutic interventions for the COVID-19 pandemic.[133] Specifically, deep learning approaches can minimize the translational gap between preclinical testing results and clinical outcomes, which is a significant problem in the rapid development of efficient treatment strategies for the COVID-19 pandemic. From a translational perspective, if broadly applied, the deep learning tools discussed here could prove helpful in developing effective treatment strategies for other complex human diseases as well, including further pandemics and other emerging infectious diseases.

**Terminology box:**

**Drug repurposing**[8]: A strategy for identifying new therapies from approved or clinically investigational drugs that have not been originally approved (also known as drug repositioning, re-tasking or re-profiling).

**Deep learning**[12]: An artificial intelligence function that mimics the workings of the human brain in processing unstructured data through many layers of neural networks.

**Machine learning**[10]: A branch of artificial intelligence in which a computer generates rules underlying or based on raw data that has been fed into it.

**Feature engineering**[11]: Feature engineering is the process of using domain knowledge of the data to create features that make machine learning algorithms work.

**Representation learning**[60]: Learning representations of the data that make it easier to extract useful information when building classifiers or other predictors.

**One-hot representation**[67]: One-hot encoding is used to represent the categorical variables as binary vectors. Each integer value is represented as a binary vector, except for the index of the integer marked by 1, all remaining values are zero.

**SMILES**[62]: Simplified Molecular Input Line Entry System (SMILES) is a linear symbol for the input and represents the molecular reactions by ASCII encoding.

**Natural Language Processing (NLP)**[64]: NLP is to process, understand and use human language (e.g., Chinese and English) by computers. It is a branch of artificial intelligence, an interdisciplinary discipline of computer science and linguistics, and is often referred to as computational linguistics (also termed computational linguistics).

**Attention mechanism**[96]: An information filtering or retrieval mechanism, similar to memory or gating, used to filter and update information.

**Deep learning Architecture:**

The fully connected **deep neural network** (DNN)[12] is the most common deep learning model. A DNN contains multiple hidden layers and each layer comprises hundreds of nonlinear process units. DNNs use multiple layers to progressively extract higher-level features from the raw input.

**Convolutional neural network** (CNN)[15] is a feed-forward neural network, which usually contains several convolution layers and subsampling layers. The parameters in convolution layers are composed of a set of filters (kernels), and the main purpose is to extract different features of the input data. The subsampling (pooling) layer is responsible for progressively reducing the spatial size of the features which decreases the number of parameters and calculations.

**Recurrent Neural Network** (RNN)[12] is a type of Neural Network where the output from the previous step is fed as input to the current step. The main and most important feature of RNN is the hidden state, which remembers some information about a sequence.

**Graph Convolutional Network** (GCN)[83] is an approach for semi-supervised learning on graph-structured data. The choice of convolutional architecture is motivated via a localized first-order approximation of spectral graph convolutions.

**Network embedding** (NE)[84] or knowledge graph embedding (KGE) also known as network representation learning aims to represent the nodes or links in a network in low-dimensional and dense vector form, so that it can have the ability of representation and reasoning in vector space.

An **autoencoder** (AE)[88] is an unsupervised learning technique for neural networks. Using backpropagation, the unsupervised algorithm continuously trains itself by setting the target output values to equal the inputs. This forces the smaller hidden encoding layer to use dimensional reduction to eliminate noise and reconstruct the inputs.

**Transformer**[96] is an architecture for transforming one sequence into another one with the help of two parts (Encoder and Decoder). The encoder consists of a set of encoding layers that processes the input iteratively one layer after another and the decoder consists of a set of decoding layers that does the same thing to the output of the encoder.

## CONCLUDING REMARKS AND FUTURE CHALLENGES

Deep learning has been widely used as a useful tool in multiple biomedical research communities, including drug repurposing. Different from physical models that depend on explicit physical equations, deep learning methods are more efficient to handle big datasets without the need for extensive computational resources, via designing pattern recognition algorithms to map the mathematical relationships between empirical observations of small molecules. Deep learning utilizes deep and specialized architectures to learn useful features from raw data. In comparison to traditional machine learning methods that rely on molecular descriptors manually constructed by domain knowledge, deep learning can automatically learn from the simple input and extract the task-specific representations of chemical structures. However, the limitations of deep learning methods lie in the requirement of large-scale, high-quality datasets for model training and the interpretability for revealing the biological significance behind the prediction. Although traditional machine learning methods can be used to solve the specific task well in some fields, with the explosive growth of data and the successful landing of AlphaFold2,[75] it is reasonable to believe that deep learning will bring the milestone to drug repurposing in the near future.

A major challenge of deep learning methods for drug repurposing is the data quality. Deep learning methods require large databases for model training. While biomedical data actually tends to be uncertain due to higher noise, incompleteness and inaccuracy. Moreover, manual annotation by experts is expensive, slow and insufficient to fill the gap between the well-labeled and unlabeled biological data. Therefore, the community effort may be a potential solution to increase the reuse and extension of compound bioactivity data. For example, an open-data web platform, termed Drug Target Commons,[134] is developed to further extract higher value from the existing and newly generated compound-protein profiling data. On the other hand, advanced deep learning algorithms specifically designed to handle such problems have gradually received much attention. For example, to handle relatively heterogeneous and scarce data, transfer learning[135] can learn a separate task from a small amount of data by using the generalizable knowledge that exist in other related tasks, which has been successfully applied in drug discovery.[136]-[138] Moreover, active learning has also been successfully applied to drug discovery.[139] Specifically, it iteratively queries the most important unlabeled samples, and then labels the samples for the next round of training to guide the improvement of the model. In addition, semi-supervised few-shot learning with better generalization can learn the limited number of cancer genomic data.[140] Another trend is precision medicine drug repurposing. Large-scale omics data, including genetics, genomics, transcriptomics, proteomics, and metabolomics, generated from cells or tissues from patient samples or disease models, will offer powerful data resources for patient stratification and to identify subtype-specific repurposable drugs for precision medicine, using deep learning approaches.[122], [141] As in global emergencies like the COVID-19 pandemic, accelerating the data sharing and collaboration of global communities will be beneficial for future development in the field of deep learning-based drug repurposing. The challenges of many DREAM communities have proved that the contribution of data usage sometimes outweighs the contribution of models in terms of prediction accuracy.[142] Overall, success in solving *in silico* drug repurposing challenges, depends on collaborative efforts among chemists, pharmacologists, data scientists, computer scientists, and drug discovery experts in improving data quality and open data sharing. Without high-quality data, even the most skilled computer science teams cannot solve the challenge with cutting-edge methods.

The most important challenges of novel deep learning methods are still the interpretability, especially in drug repurposing. Due to the complexity of the deep neural networks, it always suffers from providing the biological interpretability for the drug discovery and development communities.[143] In the field of bioinformatics and health-related, it is of importance to assess the model performance and to better understand the underlying mechanisms by interpretability.[144] The design of subtle architectures must allow for interpreting or visualizing complex relationships, which are also regarded as a challenge and opportunity for deep learning in drug repurposing. One potential direction is to adopt attention mechanisms,[96] where the model coefficients can infer the relative "importance" of each feature, and another direction may be the visualization of the network or its internal mechanism to provide interpretability.[145] Further explorations are needed to transform the "black boxes" of deep learning into "white boxes" that can be explained from a biological perspective meaningfully.

Deep learning is a promising wave for the upcoming big data-driven pharmaceutical research and drug discovery, especially in drug repurposing. While the progress of deep learning in drug repurposing is accelerating, using deep learning in clinical trials is yet to be demonstrated. People may ask: is deep learning superior to other machine learning methods for drug repurposing? We believe it is still too early to draw any firm conclusion. For tasks with structured input descriptors, deep learning seems to perform at least on par with other methods. Thus, it is better not to put all eggs in one basket. Instead, we need to fully investigate the advantages and limitations of deep learning techniques. In practice, the method used in drug repurposing might depend on which method the modeler is most familiar with and the specific problem being addressed.

**ACKNOWLEDGMENTS**

This project has been funded in whole or in part with federal funds from the National Cancer Institute, National Institutes of Health, under contract HHSN261201500003I. This work is also supported in part by NSF under grants III-1763325, III-1909323, III-2106758, and SaTC-1930941. The content of this publication does not necessarily reflect the views or policies of the Department of Health and Human Services, nor does mention of trade names, commercial products or organizations imply endorsement by the US Government. This study was supported in part by the Intramural Research Program of the NIH, National Cancer Institute, Center for Cancer Research, and the Intramural Research Program of the NIH Clinical Center.